\pdfoutput=1

\documentclass[aps,twocolumn,amsmath,amssymb,preprintnumbers,floatfix,prb,superscriptaddress,longbibliography]{revtex4-2}%{revtex4}%

\usepackage[utf8]{inputenc}
\usepackage{newtxtext}
\usepackage[upint]{newtxmath}
\usepackage{microtype}
\usepackage{textcomp}
\usepackage{eucal}
\usepackage{bm}
\usepackage{siunitx}
\usepackage{comment}
\usepackage{lipsum}

\usepackage{enumerate}
\usepackage{amsfonts}
\usepackage{amsmath}
\usepackage{amssymb}
\usepackage{color}
\usepackage{soul}

\usepackage{graphicx}

\usepackage[colorlinks,allcolors=blue]{hyperref}
\usepackage[capitalize]{cleveref}

\definecolor{DarkRed}{rgb}{0.65,0,0}%
\definecolor{Green}{rgb}{0,0.3,0.3}
\definecolor{Purple}{rgb}{0.3,0,0.65}
\definecolor{Red}{rgb}{1,0,0}
\definecolor{Blue}{rgb}{0,0,0.85}
\definecolor{Magenta}{rgb}{1,0,1}

\usepackage{amsmath} % or simply amstext

\newcommand{\Chi}[1]{\textcolor{Magenta}{{#1}}}

\newcommand{\Imag}{{\Im\mathrm{m}}}   % Imaginary part 
\newcommand{\Real}{{\mathrm{Re}}}   % Real part
\newcommand{\ve}[1]{\boldsymbol{#1}}
\newcommand{\vk}{{\ve{k}}} % Vector k
\newcommand{\abs}[1]{|#1|}

\usepackage[scale=0.94]{tgbonum}
\usepackage[mode=text]{siunitx}

\makeatletter
\DeclareTextCompositeCommand{\r}{OT1}{A}{%
  \leavevmode\vbox{%
    \offinterlineskip
    \ialign{\hfil##\hfil\cr\char23\cr\noalign{\kern-1.15ex}A\cr}%
  }%
}
\makeatother

\newcommand{\be}{\begin{equation}}
\newcommand{\ee}{\end{equation}}

\newcommand{\prlsection}[1]{\textit{#1}.\kern0.05em---\kern0.05em\ignorespaces}

\begin{document}
\title{Spin-pumping from a ferromagnetic insulator to an unconventional superconductor \\with interfacial Andreev bound-states}
\author{Chi Sun}
\affiliation{Center for Quantum Spintronics, Department of Physics, Norwegian \\ University of Science and Technology, NO-7491 Trondheim, Norway}
\author{Jacob Linder}
%\email[Corresponding author: ]{jacob.linder@ntnu.no}
\affiliation{Center for Quantum Spintronics, Department of Physics, Norwegian \\ University of Science and Technology, NO-7491 Trondheim, Norway}

\begin{abstract}
Spin-pumping from a ferromagnetic insulator into a high-$T_c$ superconductor with a $d$-wave superconducting order parameter has recently been experimentally observed. Such unconventional superconducting order is known to produce interfacial bound-states for certain crystallographic orientations. Here, we present a methodology which can be used to study spin-pumping into unconventional superconductors, including the role of interfacial bound-states. As an example, we determine how the crystallographic orientation of the $d$-wave order parameter relative the interface changes the spin-pumping effect. We find that the spin-pumping effect is slightly enhanced at low temperatures for orientations hosting interfacial bound-states compared to other superconducting states. However, the spin-pumping effect does not show a coherence peak close to $T_c$ for such orientations, and instead remains smaller than the normal state value for all $T$. For orientations not hosting interfacial bound-states, we find that the pumped spin current can be increased to several times the normal-state spin current at frequencies that are small compared to the superconducting gap. Our results show that the spin-pumping dependency on frequency and temperature changes qualitatively depending on the crystallographic orientation of unconventional superconducting order parameters relative the interface.
\end{abstract}
\maketitle

\section{Introduction}\label{sec:intro}

Spin-pumping \cite{tserkovnyak_prl_02, takahashi_handbook} is an effect where a spin current is injected from a ferromagnet with a precessing magnetization into a material in contact with the ferromagnet. Its reciprocal effect is spin-transfer torque \cite{ralph_jmmm_08}, and the two constitute central phenomena in the field of spintronics \cite{zutic_rmp_04}. 

The precessional motion of the magnetization is typically induced by an external static magnetic field that is misaligned with the magnetization. The motion is then maintained by a transverse microwave field which supplies the required energy to maintain the precession despite damping in the ferromagnet. The damping originates from a spin-lattice coupling, which ultimately causes conversion of energy to dissipated heat. The spin-pumping effect is experimentally manifested as an increase in the damping due to the angular momentum loss of the ferromagnet in the form of a spin current. The damping is measurable as an increase in the linewidth of a ferromagnetic resonance (FMR) \cite{kittel_pr_48} signal.

Spin-pumping by both ferromagnetic metals and ferromagnetic insulators (FIs) into normal metals has been studied widely in the literature \cite{tserkovnyak_rmp_05}. The injected spin current increases with the frequency of the magnetization precession, and creates a spin accumulation in the normal metal that eventually decays as one moves away from the interface with the magnetic material. When the normal metal is replaced by a superconductor, the spin-pumping effect is altered considerably \cite{bell_prl_08, morten_epl_08, jeon_natmat_18, jeon_prapp_18, yao_prb_18, umeda_apl_18}. Considering a conventional superconductor such as Al or Pb, technically denoted a BCS $s$-wave superconductor, the gap in the quasiparticle energy spectrum suppresses the spin-pumping effect at low temperatures. However, when the temperature is close to the superconducting critical temperature $T_c$, the spin pumping effect at low frequencies can be strongly enhanced in the superconducting state \cite{inoue_prb_17, kato_prb_19} compared to the normal state due to the large quasiparticle density of states close to the gap edge of the superconductor.

Spin-pumping is also possible via antiferromagnetic materials and has been considered for antiferromagnets coupled with normal metals \cite{cheng_prl_14, vaidya_science_20} and, more recently, superconductors \cite{fyhn_prb_21}. The spin-pumping effect has similar features as in the ferromagnetic case, with the exception that the FMR frequency which matches the natural precession frequency of antiferromagnetic materials is typically orders of magnitude larger than in ferromagnetic materials. This difference stems from the very large exchange coupling between the magnetic sublattices in an antiferromagnet.

Recently, spin-pumping has also been considered using unconventional superconductors \cite{ominato_prb_22, ominato_prb_22a}, such as high-$T_c$ cuprates. The superconducting order parameter in these materials can have a $d$-wave symmetry, meaning that it is anisotropic in momentum space. Quasiparticles moving in different directions in the lattice can thus experience both a different magnitude and different sign of the superconducting order parameter. Experimentally, spin-pumping in a ferromagnetic metal/$d$-wave superconductor bilayer was recently reported \cite{carreira_prb_21}. The modified damping was measured in YBa$_2$Cu$_3$O$_{7-d}$/Ni$_{80}$Fe$_{20}$ (YBCO/Py) structures, where one found a suppression of the damping upon cooling the temperature sufficiently to transition into the superconducting state of YBCO. When the temperature was lowered further, some samples showed an upturn in the linewidth, corresponding to an increase in the damping and thus more efficiently injected spin current in YBCO. This was interpreted as a consequence of the crystallographic dependence of the $d$-wave superconducting order parameter relative the Py interface: some orientations feature nodes in the order parameter, where the quasiparticle do not have any gap in their excitation spectrum. This should give rise to spin-pumping features closer to the normal metal case. Importantly, some orientations also permit for the existence of interfacial Andreev bound-states \cite{tanaka_prl_95, hu_prl_94}. These states exist at the normal-state Fermi level (zero energy) when the $d$-wave superconductor is interfaced with a non-magnetic material, like a normal metal. However, when the superconductor (SC) is interfaced with a ferromagnet, it has been shown that the bound-state energies are shifted to finite energies and that their density of states is strongly suppressed \cite{kashiwaya_prb_99}. Therefore, the low-temperature behavior of the spin-pumping in a FI/$d$-wave SC cannot immediately be explained by the formation of a large density of zero-energy interfacial Andreev bound-states that occur when a $d$-wave SC is in contact with a non-magnetic barrier. 

The spin-pumping in FI/$d$-wave SC was considered theoretically \cite{ominato_prb_22} for a fixed crystallographic orientation that does not feature any interfacial Andreev bound-states. This effect was also studied in $p$-wave superconductors, but in neither case was the effect of interfacial states taken into account \cite{ominato_prb_22a}. Here, we develop a theory for spin-pumping into unconventional superconductors for arbitrary crystallographic orientations, which thus includes the effect of interfacial bound-states. In particular, we consider a $d$-wave superconducting order parameter $\Delta = \Delta_0(T)\cos(2\theta - 2\alpha)$ where $\Delta_0(T)$ is the temperature-dependent magnitude of the superconducting order parameter, $\theta$ is the angle describing the quasiparticle motion, and $\alpha$ is a parameter that describes different crystallographic orientations of the superconductor lattice relative the interface to the FI. For $\alpha=0$, the order parameter has a $d_{x^2-y^2}$ wave form relative the interface, so that quasiparticles normally incident toward the FI experience the maximum value of the gap. For $\alpha=\pi/4$, the order parameter has a $d_{xy}$ wave form relative the interface, so that quasiparticles normally incident toward the FI experience a node of the gap. For $\alpha=0$, our results are qualitatively consistent with the recent predictions of Ref. \cite{ominato_prb_22}, although some quantitative differences are present due to the fact that we consider a specularly reflecting (clean) interface rather than a rough interface. At very small frequencies compared to the superconducting gap, the spin current is enhanced in the $d$-wave $\alpha=0$ case compared to the normal state. For $\alpha=\pi/4$, we find that the spin-pumping is different compared to both the $s$-wave and $d$-wave $\alpha=0$ cases. We find that at low temperatures and small frequencies, the spin current is slightly enhanced compared to the other superconducting pairing symmetries. Moreover, we find that the coherence peak in the spin current close to $T_c$ vanishes in the $\alpha=\pi/4$ case, so that the spin current is always smaller than that in the normal state of the system. We discuss the physical origin of all our findings. Our theory can be applied to other unconventional superconducting materials and accounts for the presence of interfacial Andreev bound-states, making it a useful tool to explore spin transport in unconventional superconducting junctions \cite{eremin_prb_06, gronsleth_prl_06, brydon_prb_09, tanaka_prb_09, takashima_prb_17, johnsen_prl_21}.

%\begin{figure}[h!]
    %\centering
    %\includegraphics[width=\columnwidth]{model.pdf}
    %\caption{(Color online) ...
		%}
    %\label{fig:model}
%\end{figure}

\section{Theory}\label{sec:theory}
Below, we present the methodology used to compute the pumped spin current. It is based on a BTK (Blonder-Tinkham-Klapwijk)-like \cite{btk} scattering approach, but modified to account for a rotating magnetization which gives rise to additional time-dependent factors in the wavefunctions. We provide detailed expressions for the wavefunctions that participate in the scattering processes, the relevant boundary conditions at the interface between the FI and the SC, and explain how the pumped spin current can be computed. The FI region occupies $x<0$ while the SC occupies $x>0$.
\text{ }\\

\subsection{Expressions in the SC}
Based on the BTK theory \cite{btk}, the propagation of quasiparticles [electron (hole)-like quasiparticles with spin up and down] in the SC can be described by the Bogoliubov-de Gennes (BdG) equation 
\begin{equation}
    \hat{H}_{S}\Psi_S = E_s \Psi_S,
\end{equation}
where $\Psi_S$ is the eigenstate with energy eigenvalue $E_s$. The Hamiltonian of the SC is given by
\begin{equation}
    \hat{H}_{S}=\begin{pmatrix}
 -{\frac{\hbar^2\triangledown^2}{2 m_e}}-\mu&0&0&\Delta\\0& -{\frac{\hbar^2\triangledown^2}{2 m_e}}-\mu&-\Delta&0\\0&-\Delta^*&{\frac{\hbar^2\triangledown^2}{2 m_e}}+\mu&0\\ \Delta^*&0&0&{\frac{\hbar^2\triangledown^2}{2 m_e}}+\mu
\end{pmatrix},
\end{equation}
where $m_e$ is the effective mass of the quasiparticles and $\mu$ is the chemical potential. The superconducting gap is denoted as $\Delta = \Delta_0(T)g(\theta)$, where $\Delta_0(T)$ is the temperature-dependent gap amplitude and $g(\theta)$ describes the superconducting pair symmetry. In this work, the conventional BCS temperature dependence $\Delta_0(T) = \Delta_0\tanh(1.74\sqrt{T_c/T-1})$ is considered. By solving the BdG equation as an eigenvalue problem, the wavefunction in the SC can be obtained.

In a $s$-wave superconductor, the superconducting gap is isotropic, i.e., $\Delta = \Delta_0(T)$ with $g(\theta)=1$. Let us first consider the wavefunction of an electron-like quasiparticle with spin up that is incoming towards the interface, which has the form of
\begin{widetext}
\begin{equation}
\Psi_S=\left[\begin{pmatrix}
    u_0\\0\\0\\v_0
\end{pmatrix}e^{-ik_{x+}x}+a_1
\begin{pmatrix}
 u_0\\0\\0\\v_0   
\end{pmatrix}e^{ik_{x+}x} + b_1\begin{pmatrix}
    v_0\\0\\0\\u_0
\end{pmatrix}e^{-ik_{x-}x} \right]e^{-\frac{iE_st}{\hbar}}e^{ik_yy}+ \left[c_1\begin{pmatrix}
    0\\u^{'}_{0}\\-v^{'}_{0}\\0
\end{pmatrix}e^{ik_{x+}^{'}x} + d_1\begin{pmatrix}
    0\\v^{'}_{0}\\-u^{'}_{0}\\0
\end{pmatrix}e^{-ik_{x-}^{'}x}\right]e^{-\frac{iE_{s}^{'}t}{\hbar}}e^{ik_yy},
\label{eq:s wave function}
\end{equation}
where $a_1$, $b_1$, $c_1$ and $d_1$ are coefficients describing normal reflection without spin-flip, Andreev reflection without spin-flip, normal reflection with spin-flip, and Andreev reflection with spin-flip, respectively. These coefficients can be determined by applying appropriate boundary conditions, which we will get back to. To differentiate it from the incident energy $E_s$, the energy after the spin-flip is denoted as $E_s^{'}$. In terms of $E_s$, we have the usual coherence factors $u_0=\sqrt{\frac{1}{2}(1+\frac{\sqrt{E_s^2-\Delta_0^2(T)}}{E_s})}$ and $v_0=\sqrt{\frac{1}{2}(1-\frac{\sqrt{E_s^2-\Delta_0^2(T)}}{E_s})}$. The $x$ components of the wave vectors are 
\begin{align}
k_{x+}&=\frac{\sqrt{2m_e(\mu+\sqrt{E_s^2-\Delta_0^2(T)})-\hbar^2k_y^2}}{\hbar},\notag\\
k_{x-}&=\frac{\sqrt{2m_e(\mu-\sqrt{E_s^2-\Delta_0^2(T)})-\hbar^2k_y^2}}{\hbar},
\end{align}
which describe the electron-like and hole-like quasiparticles, respectively. Similarly, $u_0^{'}$, $v_0^{'}$, $k_{x+}^{'}$ and $k_{x-}^{'}$  have the same forms with respect to $E_s^{'}$. Here we consider the transverse component of the wave vector $k_y=k_{\text{int}}\sin{\theta}$ is conserved across the interface, where $k_{\text{int}}$ is the corresponding normally-incident wave vector, i.e., $k_{\text{int}} = k_{x+}|_{k_y=0}=\sqrt{2m_e(\mu+\sqrt{E_s^2-\Delta_0^2(T)})}/\hbar$. 

In a $d$-wave superconductor, the superconducting gap is anisotropic, i.e., $\Delta = \Delta_0(T)g(\theta)$ with $g(\theta)=\cos{(2\theta-2\alpha)}$, as introduced in Sec. \ref{sec:intro}. The wavefunction due to an incident electron-like quasiparticle with spin up becomes
\begin{equation}
\begin{aligned}
\Psi_S &=\left[\begin{pmatrix}
    u_-\\0\\0\\v_-e^{-i\gamma_-}
\end{pmatrix}e^{-ik_{x+,-}x}+a_1
\begin{pmatrix}
 u_+\\0\\0\\v_+e^{-i\gamma_+} 
\end{pmatrix}e^{ik_{x+,+}x} + b_1\begin{pmatrix}
    v_-e^{i\gamma_-}\\0\\0\\u_-
\end{pmatrix}e^{-ik_{x-,-}x} \right]e^{-\frac{iE_st}{\hbar}}e^{ik_yy}\\
&+ \left[c_1\begin{pmatrix}
    0\\u^{'}_{+}\\-v^{'}_{+}e^{-i\gamma_+}\\0
\end{pmatrix}e^{ik_{x+,+}^{'}x} + d_1\begin{pmatrix}
0\\v^{'}_{-}e^{i\gamma_-}\\-u^{'}_{-}\\0
\end{pmatrix}e^{-ik_{x-,-}^{'}x}\right]e^{-\frac{iE_{s}^{'}t}{\hbar}}e^{ik_yy},
\end{aligned}
\label{eq:d wave function}
\end{equation}
\end{widetext}
in which the coherence factors and wave vectors are modified compared with those in the $s$-wave case due to the anisotropy. Depending on the quasiparticle motion, the coherence factors are $u_{\pm}=\sqrt{\frac{1}{2}(1+\frac{\sqrt{E_s^2-\Delta_0^2(T) g^2(\theta_{\pm})}}{E_s})}$ and $v_{\pm}=\sqrt{\frac{1}{2}(1-\frac{\sqrt{E_s^2-\Delta_0^2(T) g^2(\theta_{\pm})}}{E_s})}$ with $\theta_+ = \theta$ and $\theta_- = \pi - \theta$. Similarly, the wave vectors become \begin{align}
k_{x+,\pm}=\frac{\sqrt{2m_e(\mu+\sqrt{E_s^2-\Delta_0^2(T) g^2(\theta_{\pm})})-\hbar^2k_y^2}}{\hbar},\notag\\
k_{x-,\pm}=\frac{\sqrt{2m_e(\mu-\sqrt{E_s^2-\Delta_0^2(T) g^2(\theta_{\pm})})-\hbar^2k_y^2}}{\hbar},
\end{align}
in which the second $\pm$ sign in the subscript represents the propagation of the quasiparticles along the $\pm x$ axis. In addition, the factor $e^{i\gamma_{\pm}} =\frac{g(\theta_{\pm})}{|g(\theta_{\pm})|}$ is introduced. Here $k_y=k_{\text{int}}\sin{\theta}$ is conserved across the interface with $k_{\text{int}} = k_{x+,-}|_{k_y=0}=\sqrt{2m_e(\mu+\sqrt{E_s^2-\Delta_0^2(T)g^2(\theta_-)})}/\hbar$.

Note that the energy-dependent wave-vectors and coherence factors appearing in Eqs. (\ref{eq:s wave function}) and (\ref{eq:d wave function}) are only applicable for positive energies, i.e., $E_s>0$ and $E_s^{'}>0$. When $E_s<0$, the following replacements should be made: $k_{x\pm} \rightarrow k_{x \mp}$, $u_0 \rightarrow -v_0^*$ and $v_0 \rightarrow u_0^*$ for $s$-wave and $k_{x\pm,\pm} \rightarrow k_{x \mp, \pm}$, $u_\pm \rightarrow -v_\pm^*$ and $v_\pm \rightarrow u_\pm^*$ for $d$-wave. Similar modifications apply for $E_s^{'}<0$. A detailed explanation regarding the negative energy wavefunctions can be found in the Appendix.

\subsection{Expressions in the FI}
In the FI, the Hamiltonian for electron-like and hole-like quasiparticles has the form
\begin{equation}
    \hat{H}_{F}=
    \begin{pmatrix}
 \underline{H}_{F}&0\\0&-\underline{H}_{F}^*
\end{pmatrix}
\end{equation}
with
\begin{equation}
    \underline{H}_{F} =  -{\frac{\triangledown^2}{2 m_e}} + U + J\underline{\sigma}\cdot \textbf{M}(t),
\end{equation}
in which $\underline{\boldsymbol{\sigma}}$ denotes the Pauli matrix vector and $\underline{\ldots}$ is notation for a 2$\times$2 matrix. The potential $U$ is larger than the Fermi energy $\mu$ in the nearby SC. $J$ decribes the exchange interaction in the ferromagnet between the localized spin magnetization and the itinerant electrons. The magnetization is defined as 
\begin{equation}
   \textbf{M}(t) = (m \cos\Omega t, m \sin \Omega t, \sqrt{1-m^2}),
\end{equation}
where $m$ is the magnetization amplitude and $\Omega$ denotes the FMR frequency for spin-pumping. Here the microwave energy for FMR is considered as small compared with the exchange energy, i.e., $\hbar \Omega \ll J$. By employing a wavefunction with the structure $ e^{-\frac{iEt}{\hbar}}(e^{-\frac{i \Omega t}{2}},e^{\frac{i \Omega t}{2}},e^{\frac{i \Omega t}{2}},e^{-\frac{i \Omega t}{2}})^T $ for its time-dependence , the non-stationary Schr\"odinger equation can be solved as an eigenvalue problem. The four eigenpairs are obtained as: $E_1=E_+$ with $(a_+,b_+,0,0)^T$, $E_2=E_-$ with $(a_-,b_-,0,0)^T$, $E_3=-E_+$ with $(0,0,a_+,b_+)^T$ and $E_4=-E_-$ with $(0,0,a_-,b_-)^T$. To linear order of the adiabaticity parameter $\beta=\frac{\hbar\Omega}{2J}$, the eigenenergies are given by  
\begin{equation}
    E_\pm=U+\frac{\hbar^2(k_{x}^2+k_{y}^2)}{2m_e}\pm J(1-\beta\sqrt{1-m^2}).
\end{equation} 
The corresponding eigenstates are described by the coefficients
\begin{widetext}
\begin{equation}
    a_+=-\frac{(1-\beta)(1+\sqrt{1-m^2})-m}{\sqrt{[(1-\beta)(1+\sqrt{1-m^2})-m]^2+[(1+\beta)(1-\sqrt{1-m^2})-m]^2}},
\end{equation}
\begin{equation}
    b_+=\frac{(1+\beta)(1-\sqrt{1-m^2})-m}{\sqrt{[(1-\beta)(1+\sqrt{1-m^2})-m]^2+[(1+\beta)(1-\sqrt{1-m^2})-m]^2}},
\end{equation}
\begin{equation}
    a_-=-\frac{(1+\beta)(1-\sqrt{1-m^2})+m}{\sqrt{[(1-\beta)(1+\sqrt{1-m^2})+m]^2+[(1+\beta)(1-\sqrt{1-m^2})+m]^2}},
\end{equation}
\begin{equation}
    b_-=\frac{(1-\beta)(1+\sqrt{1-m^2})+m}{\sqrt{[(1-\beta)(1+\sqrt{1-m^2})+m]^2+[(1+\beta)(1-\sqrt{1-m^2})+m]^2}}.
\end{equation}

Based on the above, the total wavefunction in the FI is constructed as 
\begin{equation}
\Psi_F=\left[m_1
\begin{pmatrix}
 a_+e^{\frac{-i\Omega t}{2}}\\b_+e^{\frac{i\Omega t}{2}}\\0\\0   
\end{pmatrix}e^{-ik_{F1}x}e^{\frac{-iE_{1}t}{\hbar}} + n_1\begin{pmatrix}
    a_-e^{\frac{-i\Omega t}{2}}\\b_-e^{\frac{i\Omega t}{2}}\\0\\0
\end{pmatrix}e^{-ik_{F2}x}e^{\frac{-iE_{2}t}{\hbar}} + p_1\begin{pmatrix}
    0\\0\\a_+e^{\frac{i\Omega t}{2}}\\b_+e^{\frac{-i\Omega t}{2}}
\end{pmatrix}e^{-ik_{F3}x}e^{\frac{-iE_{3}t}{\hbar}}
+ q_1\begin{pmatrix}
    0\\0\\a_-e^{\frac{i\Omega t}{2}}\\b_-e^{\frac{-i\Omega t}{2}}
\end{pmatrix}e^{-ik_{F4}x}e^{\frac{-iE_{4}t}{\hbar}}\right]e^{ik_yy},
\end{equation}
\end{widetext} 
where $m_1$, $n_1$, $p_1$ and $q_1$ are transmission coefficients to be determined by applying appropriate boundary conditions. In order to match the time-dependence of the wavefunction components on the FI and SC sides, we can obtain $E_s^{'}=E_s-\hbar\Omega$ and $E_1=E_2=E_3=E_4=E_s-\frac{\hbar\Omega}{2}$. In terms of $E_s$, the corresponding $x$ component of the four wave vectors in the FI are expressed as 
\begin{equation}k_{F1}=\frac{\sqrt{2m_e[E_s-U-J(1-\beta\sqrt{1-m^2}+\beta)]-\hbar^2k_y^2}}{\hbar},\end{equation}
\begin{equation}k_{F2}=\frac{\sqrt{2m_e[E_s-U+J(1-\beta\sqrt{1-m^2}-\beta)]-\hbar^2k_y^2}}{\hbar},\end{equation} 
\begin{equation} k_{F3}=\frac{\sqrt{2m_e[-E_s-U-J(1-\beta\sqrt{1-m^2}-\beta)]-\hbar^2k_y^2}}{\hbar}\end{equation}
and \begin{equation}k_{F4}=\frac{\sqrt{2m_e[-E_s-U+J(1-\beta\sqrt{1-m^2}+\beta)]-\hbar^2k_y^2}}{\hbar}.\end{equation} Note that all wave numbers in the FI possess imaginary values since a large potential $U$ is required to ensure the ferromagnet to be insulating. To ensure this, we use $U=2\mu$ throughout this work.

\subsection{Boundary conditions and spin pumping current} 
The boundary conditions at the FI/SC interface (i.e., $x=0$) are 
\begin{equation}
    \Psi_F\big|_{x=0}=\Psi_S\big|_{x=0},
\end{equation} 
\begin{equation}
    \partial_x\Psi_S\big|_{x=0}-\partial_x\Psi_F\big|_{x=0}=2Zk_F\Psi_F\big|_{x=0},
\end{equation} 
which describe the continuity of the wavefunction and its derivative appropriate for the $\delta$ function potential at the interface, respectively. $Z$ is a dimensionless parameter which characterizes the interface transparency.

\begin{figure*}[t!]
  \centering
  \includegraphics[width=1.0\linewidth]{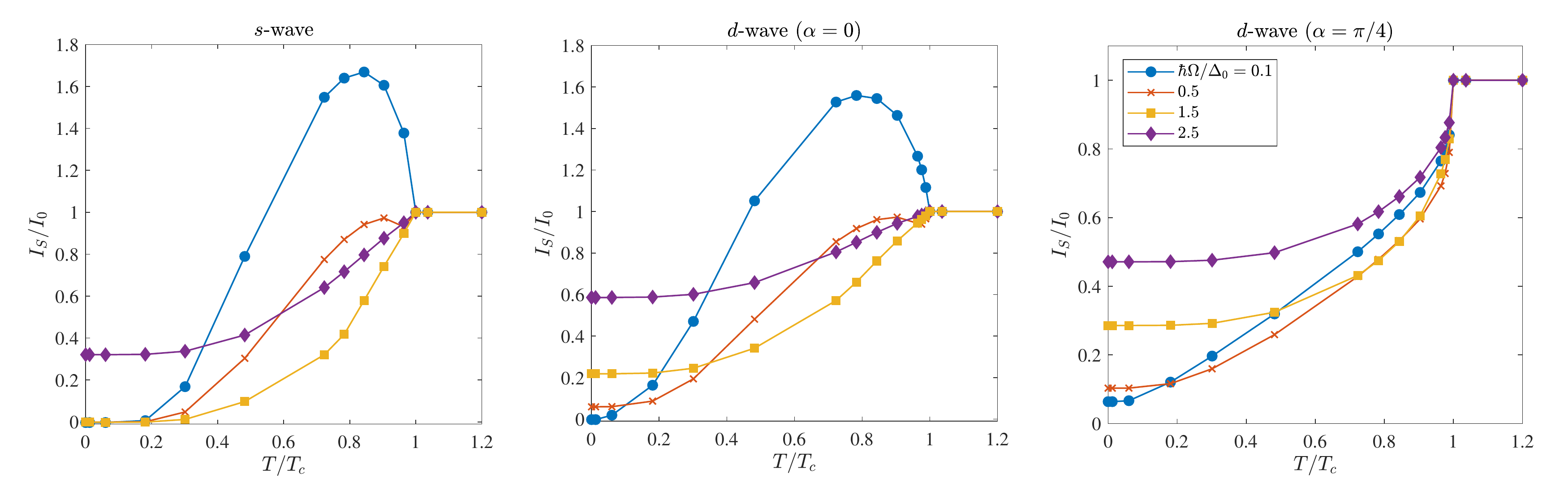}
  \caption{(Color online) Temperature-dependence of the pumped spin current into the superconductor. The spin current is normalized on its normal-state value and shown for the $s$-wave (left), $d$-wave $\alpha=0$ (middle), and $d$-wave $\alpha=\pi/4$ (right) cases for several frequencies $\Omega$. Here we use $U=2\mu$, where $\mu = \hbar^2k_F^2/2m_e$ is the Fermi energy with the Fermi wave vector $k_F=1$ \AA$^{-1}$. 
  }%
  \label{fig:temp}
\end{figure*}

The longitudinal spin-pumping current polarized along the $z$ axis can be calculated as 
\begin{equation}
    I_{s}=\int_{-W_l}^{W_h}\int_{-\pi/2}^{\pi/2} j_{sz}(E_s,\theta)f_0(E_s)N_s\cos{\theta}d\theta dE_s,
\label{eq:Is}
\end{equation}
where the quantum mechanical spin current $j_{sz}=-\frac{\hbar^2}{2m_e}\Imag[\Psi_S^{\dagger}\partial_x\mathbf{\hat{\sigma_z}}\Psi_S]$ with $\mathbf{\hat{\sigma_z}}=\begin{pmatrix}
 \underline{\sigma}_z&0\\0&\underline{\sigma}_z^*
\end{pmatrix}$ produced by a given incoming particle. $f_0(E_s)$ denotes the Fermi-Dirac distribution and $N_s=\Real[\frac{\abs{E_s}}{\sqrt{E_s^2-\Delta^2}}]$ is the density of states in the SC. Consider the specific case of an incoming spin-up electron-like quasiparticle from the SC. To ensure that the $x$-component of the wave vectors (before and after spin-flip) in the SC are real and propagating for  normal incidence (i.e., $\theta = 0$), the upper and lower integration limit for $E_s$ are determined by $W_h = \sqrt{\mu^2+\Delta_0^2(T)}$ and $W_l=\sqrt{\mu^2+\Delta_0^2(T)}-\hbar\Omega$, in which the $\hbar \Omega$ shift originates from the energy relation $E_s^{'} = E_s - \hbar \Omega$. However, in practice a much smaller integration range for $E_s$ can be employed to greatly enhance the numerical integration efficiency for the SC when $T \ll T_c$. As explained in the Appendix, a temperature-dependent integration range may be utilized in practice. In all plots, we set the barrier parameter to $Z=2$. Varying $Z$ does not give rise to any qualitatively new features, since the ferromagnetic region is insulating and thus total reflection (either normal or Andreev) of incoming particles.

Eq. (\ref{eq:Is}) covers the spin-pumping current induced by the electron-like quasiparticle with spin up (i.e., electron-up) incident. The total spin current is given by the sum of electron-up and electron-down contributions. In the Appendix, we present the wavefunctions in the SC induced by electron-down, hole-up and hole-down incidents. Based on these wavefunctions, it is shown that the spin-pumping current can also be calculated as the sum of hole-up, hole-down and additional background contributions (see Appendix for details). 

The ratio between the pumped spin currents in the superconducting and normal metal cases correspond to the ratio of the Gilbert damping corrections in these two cases. The reason is that the dc component of the pumped spin current (proportional to $\textbf{M} \times d\textbf{M}/dt$) enters the Landau-Lifshitz-Gilbert as an effective damping term \cite{tserkovnyak_rmp_05}. Therefore, it renormalizes the Gilbert damping coefficient of the material. Computing the ratio of the magnitudes of the dc component of the pumped spin current, as done here, is thus equivalent to computing the ratio of the spin pumping induced correction to the Gilbert damping in the superconducting and normal states.

We underline again that our methodology allows us to account for the presence of interface bound-states, unlike the approaches used in Refs. \cite{inoue_prb_17, kato_prb_19, ominato_prb_22, ominato_prb_22a}.

\section{Results and Discussion}\label{sec:results}

\subsection{Temperature dependence}\label{sec:temp}

In Fig. \ref{fig:temp}, we investigate the temperature dependence of  the pumped spin current into the SC. Due to the existence of interfacial Andreev bound-states in N/$d$-wave or I/$d$-wave junctions in the $\alpha=\pi/4$ case, one might expect that the spin-pumping effect would be strongly enhanced compared to the $s$-wave or $d$-wave $\alpha=0$ case. However, our results do not show any such strong enhancement. To understand these results, it is instructive to consider the quantum transport of a ferromagnet/insulator/$d$-wave superconductor junction as considered in Ref. \cite{kashiwaya_prb_99}. With increasing polarization of the ferromagnet and the interface region, the density of interfacial bound-states is reduced and shifted away from the Fermi level. Moreover, it was shown in Ref. \cite{kashiwaya_prb_99} that the interfacial states do not contribute to the spin conductance of the junction, whereas they contribute strongly to the charge conductance. This can be understood from the fact that the interfacial states are confined to a distance of order superconducting coherence length from the interface, and are thereafter converted into a Cooper pair charge supercurrent propagating into the bulk of the SC. However, the Cooper pairs in the bulk of the $d$-wave SC cannot carry a spin current since they are in a singlet state, which means that the interfacial bound-states cannot contribute to spin transport into the SC. 

\begin{figure*}[t!]
  \centering
  \includegraphics[width=1.0\linewidth]{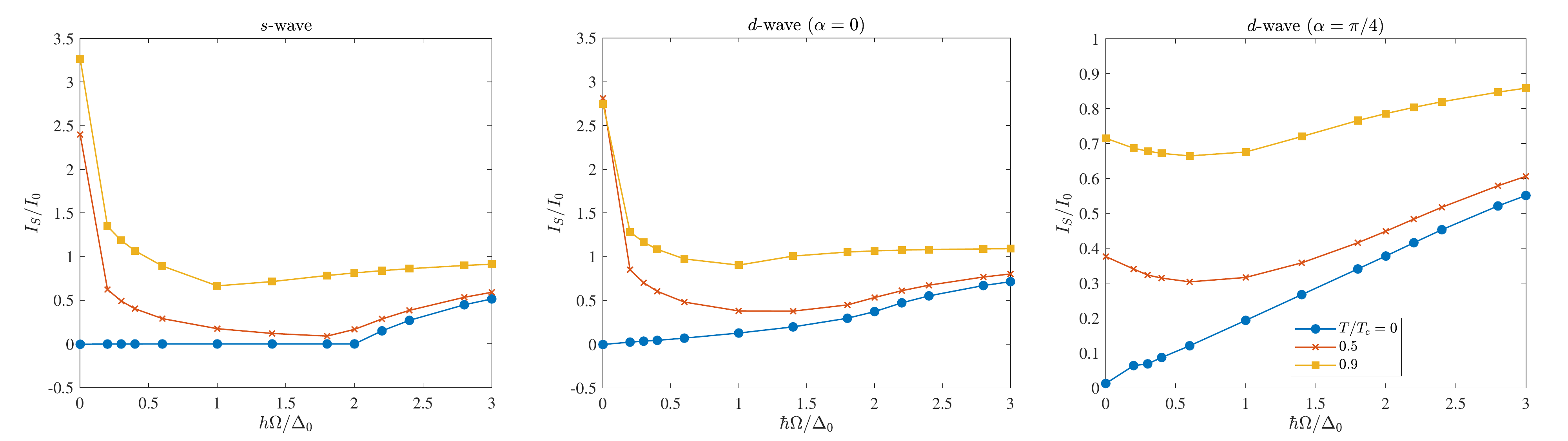}
  \caption{(Color online) Frequency-dependence of the pumped spin current into the superconductor. The spin current is normalized on its normal-state value and shown for the $s$-wave (left), $d$-wave $\alpha=0$ (middle), and $d$-wave $\alpha=\pi/4$ (right) cases for several temperatures $T$.
  }%
  \label{fig:omega}
\end{figure*}

This is consistent with our results. In the $\alpha=\pi/4$ case, there is only a slight increase in the spin-pumping current at low temperatures and small frequencies $\hbar \Omega/\Delta_0 \ll 1$ compared to the $s$-wave and $d$-wave $\alpha=0$ case. Rather than attributing this to the effect of interfacial bound-states, this small increase in current should instead stem from the nodal positioning of the gap relative the interface. Normally incident particles feel no gap, i.e. $\Delta=0$, in the $\alpha=\pi/4$ case, so that the spin current is not blocked by the excitation spectrum gap unlike the $s$-wave and $d$-wave $\alpha=0$ cases.

The second qualitative difference in the $\alpha=\pi/4$ case is the absence of the coherence peak in the spin-pumping current for temperatures $T \simeq T_c$. Such a peak appears in the $s$-wave at low frequencies $\hbar\Omega \ll \Delta_0$. The reason is that close to $T_c$, the superconducting gap is suppressed, which makes it possible for spin-flip excitations of quasiparticles with energy $\hbar\Omega$ to occur. Such excitations can carry a spin current. For energies right above the superconducting gap, the density of quasiparticle states is strongly enhanced (the so-called coherence peak). This large number of available states is the reason for the superconducting spin current exceeding the normal state spin current at low frequencies, since in the normal metal case there is a range of energies $\hbar \Omega$ around the Fermi level which contribute to the pumped spin current. As $\Omega$ increases, the relative increase in the superconducting spin current vs. normal spin current is lost since the normal state contribution grows more rapidly. A similar reasoning applies to the $d$-wave $\alpha=0$ case, although the peaked spin current close to $T_c$ is smaller than in the $s$-wave case due to the nodes in the anisotropic $d$-wave gap. As we will discuss later, this becomes different for very small frequencies compared to the superconducting gap, which are experimentally the most straightforward ones to obtain.

The difference between the two cases described above and the $d$-wave $\alpha=\pi/4$ case which causes the coherence peak at $T\simeq T_c$ to be absent, is the nodal orientation of the gap relative the interface. Quasiparticle trajectories close to normal incidence $\theta=0$ will contribute the most to spin transport. The gap is very small close to normal incidence and indeed vanishes exactly at $\theta=0$. Therefore, the coherence peak effect described for the $s$-wave case is present already at low temperatures (due to the smallness of the gap) for quasiparticle trajectories close to $\theta=0$. It is not present for trajectories away from normal incidence, since the gap eventually recovers its full value at $\theta=\pi/4$. This is the reason for why the spin current for $\alpha=\pi/4$ is slightly enhanced at low frequencies compared to the other order parameter cases. As the temperature gets closer to $T_c$, however, there is no large pile-up of available quasiparticle states close to normal incidence that suddenly become accessible and boost the superconducting spin current, unlike in the $s$-wave and $d$-wave $\alpha=0$ cases. Therefore, the spin current in the superconducting case always remains smaller than the normal state value for the $\alpha=\pi/4$ crystallographic orientation.

It is pertinent to discuss our findings in relation to other works. For instance, Ref. \cite{ahari_prb_21} found that the spin pumping current into a SC could be enhanced by bound-states present in the superconducting region. We note that this result is consistent with our result, because the enhancement in Ref. \cite{ahari_prb_21} only took place when the SC had a thickness comparable to the superconducting coherence length. For such a small SC, the interfacial states extend over the
whole superconducting region and can contribute to transport. In our
case of a bulk SC (semi-infinite), they cannot, and ultimately
lead to a suppressed spin pumping current.

The most relevant experiment to date regarding spin pumping in a ferromagnet/$d$-wave SC junction was reported in Ref. \cite{carreira_prb_21}. For some of their samples, the authors reported observation of an upturn of the Gilbert damping coefficient at low temperatures, indicating an enhanced spin pumping current. However, we note that the analysis in the experiment \cite{carreira_prb_21} was done using a model for the bound-states in the absence
of magnetism, in effect for a normal metal/$d$-wave SC junction. Since Ref. \cite{kashiwaya_prb_99} showed
that the polarization of the ferromagnet strongly influences the bound-states, and in fact
that the contribution from the Fermi level to the spin conductance of a
ferromagnet/$d$-wave SC junction is strongly suppressed, we
believe these properties should influence the interpretation of the experimental results.

\subsection{Frequency dependence}\label{sec:omega}
We now consider the dependence of the ratio between the pumped spin current in the superconducting and normal-state on frequency in Fig. \ref{fig:omega}.
At $\Omega=0$, the spin current vanishes for all systems since there is no magnetization precession and thus no pumped spin current. We note that for temperatures approaching $T_c$, there is an abrupt increase in the the ratio of the spin current in the superconducting state and normal state vs. frequency as soon as $\Omega$ becomes non-zero in the $s$-wave and $d$-wave $\alpha=0$ cases. This means that the spin-pumping is much more efficient in the superconducting state compared to the normal state. This can be understood from the argument provided in the previous section: at small frequencies $\hbar \Omega \ll \Delta_0$, the large density of quasiparticles that can carry spin current with energies close to the gap edge in the SC becomes available and exceeds the normal-state spin current for small frequencies. Interestingly, the figure shows that the spin current in the anisotropic $d$-wave $\alpha=0$ case is several times larger than the normal-state current for low frequencies $\hbar\Omega/\Delta_0 \ll 1$ at $T/T_c \simeq 0.5.$ This cannot be explained by the nodal quasiparticle transport in the $d$-wave case compared to the isotropic $s$-wave gap, since such an effect would make the spin current closer to its normal-state value. Instead, this enhancement in the $d$-wave case stems from the pileup of quasiparticle density of states near energies $E_s = \Delta_0(T)g(\theta)$. Since the gap magnitude varies with $\theta$, and is generally smaller or equal to $\Delta_0$, quasiparticles are more easily accessible via thermal excitations for small frequencies. 

In ferromagnets, the FMR resonance frequencies typically lies in the 1 GHz - 100 GHz range. A conventional BCS $s$-wave SC usually has a gap of order 1 meV, corresponding to roughly 250 GHz, whereas the gap magnitude in high-$T_c$ $d$-wave SCs can be ten times as large. It is clear that $\hbar\Omega/\Delta_0 \ll 1$ for such parameters, suggesting that the large enhancement of the spin current in the superconducting state relative the normal state that takes place exactly for $\hbar\Omega/\Delta_0 \ll 1$ is accessible experimentally. Two remarks are in order here. First, it is possible to access higher frequencies relative the gap, $\hbar\Omega/\Delta_0 \sim 1$, by considering spin-pumping into proximity-induced superconducting regions. In effect, by considering spin-pumping from a FI into a normal metal where the latter is proximitized by a SC, the normal metal may effectively be considered as a SC with a smaller gap $\Delta_\text{eff}$ compared to its host SC. Secondly, larger ratios of $\hbar\Omega/\Delta_0$ are also attainable by replacing the FI with an antiferromagnetic insulator \cite{fyhn_prb_21}. In antiferromagnets (AFMs), the FMR resonance frequency is of order THz, and recent theory suggests that the spin-pumping dependence on temperature and frequency is similar in an AFI/SC as in an FI/SC bilayer \cite{fyhn_prb_21}.

\subsection{Crystallographic axis dependence}\label{sec:alpha}

\begin{figure}[t!]
  \centering
  \includegraphics[width=1.0\linewidth]{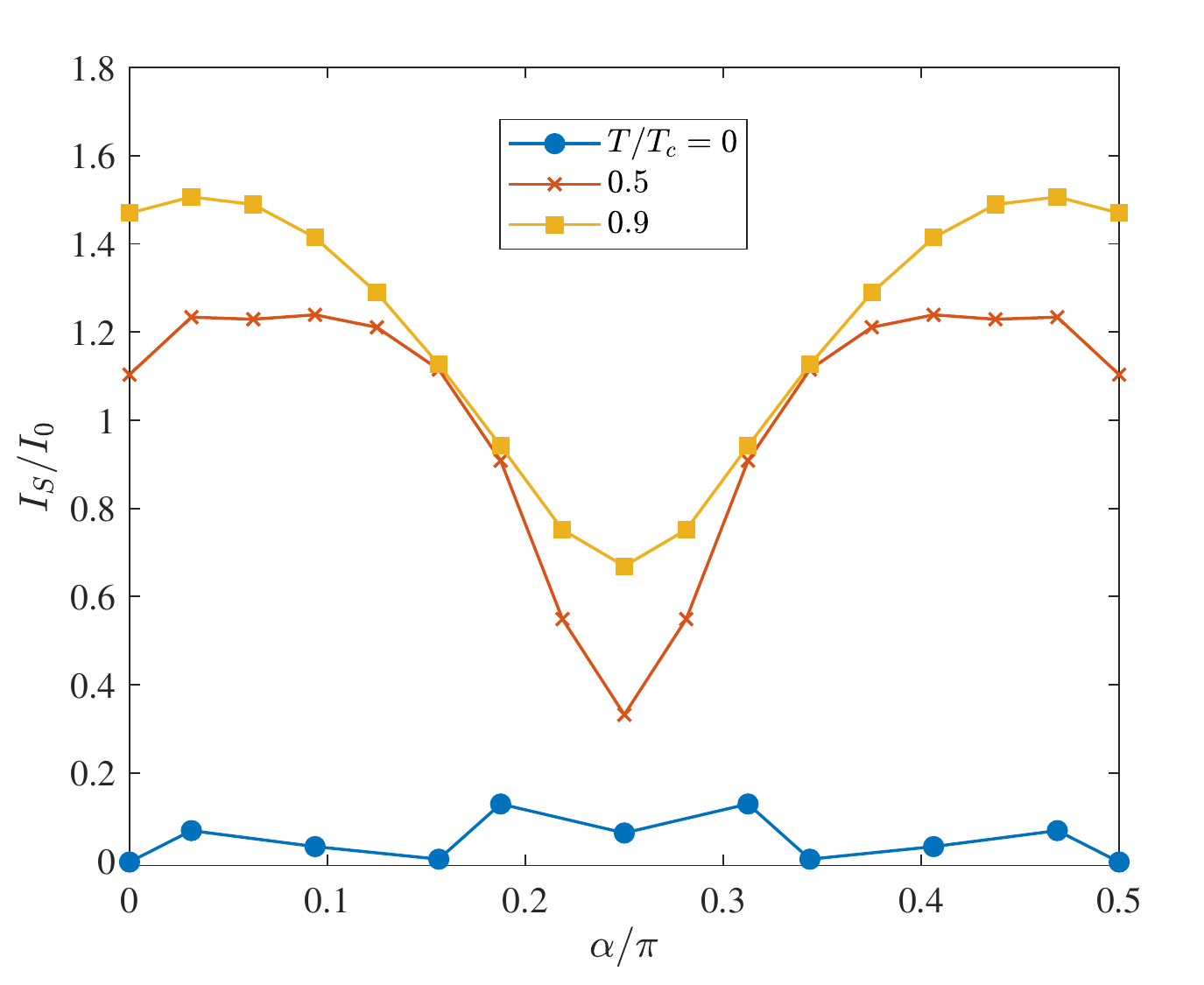}
  \caption{(Color online) Spin current in the superconducting state normalized against the normal-state value of the spin current vs. crystallographic axis orientation of the $d$-wave superconducting material relative the interface. The $\alpha=0$ case corresponds to $d_{x^2-y^2}$-wave pairing relative the interface, whereas $\alpha=\pi/4$ corresponds to $d_{xy}$-wave pairing (featuring zero-energy interface states in the absence of ferromagnetism). We have set $\hbar\Omega/\Delta_0 = 0.1.$
  }%
  \label{fig:d_alpha}
\end{figure}

In Fig. \ref{fig:d_alpha}, we consider exclusively the $d$-wave case and vary continuously the misalignment $\alpha$ between the antinodal direction of the gap and the interface normal. Experimentally, this corresponds to varying the crystallographic orientation of how the superconducting sample is grown on top of the FI. We see that the $\alpha=\pi/4$, corresponding to the orientation which hosts zero-energy interfacial states in the absence of ferromagnetism, provides the smallest spin current for moderate to high temperatures relative $T_c$. Only at low temperatures $T\ll T_c$, the spin current for $\alpha=\pi/4$ is comparable to other crystallographic orientations. This result is consistent with the explanation given in Sec. \ref{sec:temp}, which shows why the $\alpha=\pi/4$ direction does not provide any spin current enhancement relative the normal-state as the temperature approaches $T_c$. The fact that $\alpha=\pi/4$ hosts interfacial bound-states, pinned to zero-energy for any angles of incidence $\theta$ of the quasiparticles in the absence of magnetism, but shifted to finite energies in the presence of magnetism, does not help boosting the spin current. As mentioned previously, these states are confined to close to the interface region and cannot be converted into a Cooper pair current carrying the spin in the bulk of the superconductor. 

We mention in passing how impurity scattering is expected to influence the results. In the $s$-wave case, adding weak impurity scattering to the SC should not lead to any qualitative differences regarding the pumped spin current, since the superconducting order parameter is robust against impurities in the Born limit (weak impurity potential, high concentration of impurities). However, even weak impurities will have a significant impact on the pumped current in the $d$-wave case since such an anisotropic order parameter is strongly suppressed in the presence of impurity scattering.

\section{Summary}\label{sec:summary}
Motivated by a recent experiment demonstrating spin-pumping from a ferromagnetic insulator into a high-$T_c$ superconductor with a $d$-wave superconducting order parameter, we have presented a methodology which can be used to study spin-pumping into unconventional superconductors, including the role of interfacial bound-states. To be concrete, we have focused on how the crystallographic orientation of a $d$-wave superconducting order parameter relative the interface changes the spin-pumping effect. Such unconventional superconductivity is known to produce interfacial bound-states for certain crystallographic orientations. We have shown that the spin-pumping effect is slightly enhanced at low temperatures for orientations hosting interfacial bound-states in the absence of ferromagnetism compared to other superconducting states. However, the spin-pumping effect did not show a coherence peak close to $T_c$ for such orientations, and instead remains smaller than the normal state value for all $T$. For orientations not hosting interfacial bound-states, we found that the pumped spin current can be increased to several times the normal-state spin current at frequencies that are small compared to the superconducting gap. Our results show that the spin-pumping dependency on frequency and temperature changes qualitatively depending on the crystallographic orientation of unconventional superconducting order parameters relative the interface.

\begin{acknowledgments}
We thank J. A. Ouassou for useful discussions. We acknowledge funding via the Research Council of Norway Grant numbers 323766, as well as through its Centres of Excellence funding scheme QuSpin, project number 262633.
\end{acknowledgments}

\appendix

\section{Complete wavefunctions}
In the main text, we gave a detailed expression for the wavefunction for incoming  electron-like quasiparticles with spin up (referred to as electron-up) in the superconductor. Here, we provide the corresponding expressions for the other possible incoming quasiparticles. For notation simplicity, we use $e\downarrow$, $h \uparrow$ and $h \downarrow$ to represent the electron-down, hole-up and hole-down incidents, respectively. In a $s$-wave SC, the wavefunctions are 
\begin{widetext}
\begin{equation}
\Psi_{S,e\downarrow}=\left[\begin{pmatrix}
    0\\u_0\\-v_0\\0
\end{pmatrix}e^{-ik_{x+}x}+a_2
\begin{pmatrix}
 0\\u_0\\-v_0\\0   
\end{pmatrix}e^{ik_{x+}x} + b_2\begin{pmatrix}
    0\\v_0\\-u_0\\0
\end{pmatrix}e^{-ik_{x-}x} \right]e^{-\frac{iE_st}{\hbar}}e^{ik_yy}+ \left[c_2\begin{pmatrix}
    u^{'}_{0}\\0\\0\\v^{'}_{0}
\end{pmatrix}e^{ik_{x+}^{'}x} + d_2\begin{pmatrix}
    v^{'}_{0}\\0\\0\\u^{'}_{0}
\end{pmatrix}e^{-ik_{x-}^{'}x}\right]e^{-\frac{iE_{s}^{'}t}{\hbar}}e^{ik_yy},
\end{equation}
\begin{equation}
\Psi_{S,h\uparrow}=\left[\begin{pmatrix}
    0\\v_0\\-u_0\\0
\end{pmatrix}e^{ik_{x-}x}+a_3
\begin{pmatrix}
 0\\v_0\\-u_0\\0   
\end{pmatrix}e^{-ik_{x-}x} + b_3\begin{pmatrix}
    0\\u_0\\-v_0\\0
\end{pmatrix}e^{ik_{x+}x} \right]e^{-\frac{iE_st}{\hbar}}e^{ik_yy}+ \left[c_3\begin{pmatrix}
    v^{'}_{0}\\0\\0\\u^{'}_{0}
\end{pmatrix}e^{-ik_{x-}^{'}x} + d_3\begin{pmatrix}
    u^{'}_{0}\\0\\0\\v^{'}_{0}
\end{pmatrix}e^{ik_{x+}^{'}x}\right]e^{-\frac{iE_{s}^{'}t}{\hbar}}e^{ik_yy},
\end{equation}
\begin{equation}
\Psi_{S,h\downarrow}=\left[\begin{pmatrix}
    v_0\\0\\0\\u_0
\end{pmatrix}e^{ik_{x-}x}+a_4
\begin{pmatrix}
 v_0\\0\\0\\u_0   
\end{pmatrix}e^{-ik_{x-}x} + b_4\begin{pmatrix}
    u_0\\0\\0\\v_0
\end{pmatrix}e^{ik_{x+}x} \right]e^{-\frac{iE_st}{\hbar}}e^{ik_yy}+ \left[c_4\begin{pmatrix}
    0\\v^{'}_{0}\\-u^{'}_{0}\\0
\end{pmatrix}e^{-ik_{x-}^{'}x} + d_4\begin{pmatrix}
    0\\u^{'}_{0}\\-v^{'}_{0}\\0
\end{pmatrix}e^{ik_{x+}^{'}x}\right]e^{-\frac{iE_{s}^{'}t}{\hbar}}e^{ik_yy}.
\end{equation}
In a $d$-wave SC, we have
\begin{equation}
\begin{aligned}
\Psi_{S,e\downarrow} &=\left[\begin{pmatrix}
    0\\u_-\\-v_-e^{-i\gamma_-}\\0
\end{pmatrix}e^{-ik_{x+,-}x}+a_2
\begin{pmatrix}
 0\\u_+\\-v_+e^{-i\gamma_+}\\0 
\end{pmatrix}e^{ik_{x+,+}x} + b_2\begin{pmatrix}
    0\\v_-e^{i\gamma_-}\\-u_-\\0
\end{pmatrix}e^{-ik_{x-,-}x} \right]e^{-\frac{iE_st}{\hbar}}e^{ik_yy}\\
&+ \left[c_2\begin{pmatrix}
    u^{'}_{+}\\0\\0\\v^{'}_{+}e^{-i\gamma_+}
\end{pmatrix}e^{ik_{x+,+}^{'}x} + d_2\begin{pmatrix}
v^{'}_{-}e^{i\gamma_-}\\0\\0\\u^{'}_{-}
\end{pmatrix}e^{-ik_{x-,-}^{'}x}\right]e^{-\frac{iE_{s}^{'}t}{\hbar}}e^{ik_yy},
\end{aligned}
\end{equation}
\begin{equation}
\begin{aligned}
\Psi_{S,h\uparrow} &=\left[\begin{pmatrix}
    0\\v_+e^{i\gamma_+}\\-u_+\\0
\end{pmatrix}e^{ik_{x-,+}x}+a_3
\begin{pmatrix}
 0\\v_-e^{i\gamma_-}\\-u_-\\0 
\end{pmatrix}e^{-ik_{x-,-}x} + b_3\begin{pmatrix}
    0\\u_+\\-v_+e^{-i\gamma_+}\\0
\end{pmatrix}e^{ik_{x+,+}x} \right]e^{-\frac{iE_st}{\hbar}}e^{ik_yy}\\
&+ \left[c_3\begin{pmatrix}
v^{'}_{-}e^{i\gamma_-}\\0\\0\\u_-^{'}
\end{pmatrix}e^{-ik_{x-,-}^{'}x} + d_3\begin{pmatrix}
u_+^{'}\\0\\0\\v^{'}_{+}e^{-i\gamma_+}
\end{pmatrix}e^{ik_{x+,+}^{'}x}\right]e^{-\frac{iE_{s}^{'}t}{\hbar}}e^{ik_yy},
\end{aligned}
\end{equation}
\begin{equation}
\begin{aligned}
\Psi_{S,h\downarrow} &=\left[\begin{pmatrix}
    v_+e^{i\gamma_+}\\0\\0\\u_+
\end{pmatrix}e^{ik_{x-,+}x}+a_4
\begin{pmatrix}
 v_-e^{i\gamma_-}\\0\\0\\u_-
\end{pmatrix}e^{-ik_{x-,-}x} + b_4\begin{pmatrix}
    u_+\\0\\0\\v_+e^{-i\gamma_+}
\end{pmatrix}e^{ik_{x+,+}x} \right]e^{-\frac{iE_st}{\hbar}}e^{ik_yy}\\
&+ \left[c_4\begin{pmatrix}
    0\\v^{'}_{-}e^{i\gamma_-}\\-u^{'}_{-}\\0
\end{pmatrix}e^{-ik_{x-,-}^{'}x} + d_4\begin{pmatrix}
0\\u^{'}_{+}\\-v^{'}_{+}e^{-i\gamma_+}\\0
\end{pmatrix}e^{ik_{x+,+}^{'}x}\right]e^{-\frac{iE_{s}^{'}t}{\hbar}}e^{ik_yy}.
\end{aligned}
\end{equation}

Similarly, by matching the time-dependence of each component of the wavefunctions on both FI and SC sides, we can obtain that $E_s^{'}=E_s-\hbar\Omega$ for $h\downarrow$ incident but $E_s^{'}=E_s+\hbar\Omega$ for $e\downarrow$ and $h\uparrow$ incidents. Note that the above wave functions are only applicable for positive energies, i.e., $E_s>0$ and $E_s^{'}>0$. For negative energy wavefunctions, the energy-dependent wave vectors and coherence factors should be modified, as indicated in the main text. 

\section{Negative energy wavefunctions in the SC}
The wavefunctions in the SC are obtained by solving the BdG equation $\hat{H}_{S}\Psi_S = E_s \Psi_S$, in which $E_s$ can be either positive or negative. It is found that if $(\psi_1,0,0,\psi_2)^T$ is an eigenstate for $E_s = \sqrt{\xi^2 + \Delta^2}$, then $\Chi{(-\psi_2^*,0,0,\psi_1^*)^T}$ is an eigenstate for $E_s = -\sqrt{\xi^2 + \Delta^2}$ with $\xi=\frac{\hbar^2\textbf{k}^2}{2m_e}-\mu$.

Let us consider the wavefunction in a $s$-wave SC describing an electron-up incident as an example. For the positive eigenenergy $E_s = \sqrt{\xi^2 + \Delta^2}$, the corresponding solution of the eigenstate are 
\begin{equation}
    \psi_1=\sqrt{\frac{1}{2}(1+\frac{\xi}{E_s})},\  \psi_2=\sqrt{\frac{1}{2}(1-\frac{\xi}{E_s})}.
\end{equation} 
There are two possible choices of $\xi$:
\begin{equation}
  \xi = \sqrt{E_s^2-\Delta_0(T)^2}: \ \psi_1=u_0, \ \psi_2 = v_0,\ k_{x+}=\frac{\sqrt{2m_e(\mu+\sqrt{E_s^2-\Delta_0^2(T)})-\hbar^2k_y^2}}{\hbar},
\end{equation}
\begin{equation}
  \xi = -\sqrt{E_s^2-\Delta_0(T)^2}: \ \psi_1=v_0, \ \psi_2 = u_0,\ k_{x-}=\frac{\sqrt{2m_e(\mu-\sqrt{E_s^2-\Delta_0^2(T)})-\hbar^2k_y^2}}{\hbar}.
\end{equation}
On the other hand, for the negative eigenenergy $E_s = -\sqrt{\xi^2 + \Delta^2}$, the corresponding solution of the eigenstate changes to 
\begin{equation}
    \psi_1=-\sqrt{\frac{1}{2}(1+\frac{\xi}{E_s})}^*,\  \psi_2=\sqrt{\frac{1}{2}(1-\frac{\xi}{E_s})}^*.
\end{equation} 
As before, there are two possible choices of $\xi$. We have
\begin{equation}
  \xi = -\sqrt{E_s^2-\Delta_0(T)^2}: \ \psi_1=-v_0^*, \ \psi_2 = u_0^*,\ k_{x-}=\frac{\sqrt{2m_e(\mu-\sqrt{E_s^2-\Delta_0^2(T)})-\hbar^2k_y^2}}{\hbar},
\end{equation}
\begin{equation}
  \xi = \sqrt{E_s^2-\Delta_0(T)^2}: \ \psi_1=-u_0^*, \ \psi_2 = v_0^*,\ k_{x+}=\frac{\sqrt{2m_e(\mu+\sqrt{E_s^2-\Delta_0^2(T)})-\hbar^2k_y^2}}{\hbar}.
\end{equation}
Based on the above, the negative energy version of Eq. (\ref{eq:s wave function}) in the main text is modified to
\begin{equation}
\Psi_S=\left[\begin{pmatrix}
    -v_0^*\\0\\0\\u_0^*
\end{pmatrix}e^{-ik_{x-}x}+a_1
\begin{pmatrix}
 -v_0^*\\0\\0\\u_0^*   
\end{pmatrix}e^{ik_{x-}x} + b_1\begin{pmatrix}
    u_0^*\\0\\0\\-v_0^*
\end{pmatrix}e^{-ik_{x+}x} \right]e^{-\frac{iE_st}{\hbar}}e^{ik_yy}+ \left[c_1\begin{pmatrix}
    0\\-v^{'*}_{0}\\-u^{'*}_{0}\\0
\end{pmatrix}e^{ik_{x-}^{'}x} + d_1\begin{pmatrix}
    0\\u^{'*}_{0}\\v^{'*}_{0}\\0
\end{pmatrix}e^{-ik_{x+}^{'}x}\right]e^{-\frac{iE_{s}^{'}t}{\hbar}}e^{ik_yy}
\end{equation}
when both $E_s$ and $E_s^{'}$ are negative. If either $E_s$ or $E_s^{'}$ is negative, only the wave vectors and coherence factors regarding the negative energy should be modified. A similar procedure can be applied to the $d$-wave SC and the corresponding modifications for the negative energy wavefunctions are indicated in the main text.

\end{widetext}

\begin{figure*}[t!]
  \centering
  \includegraphics[width=1.0\linewidth]{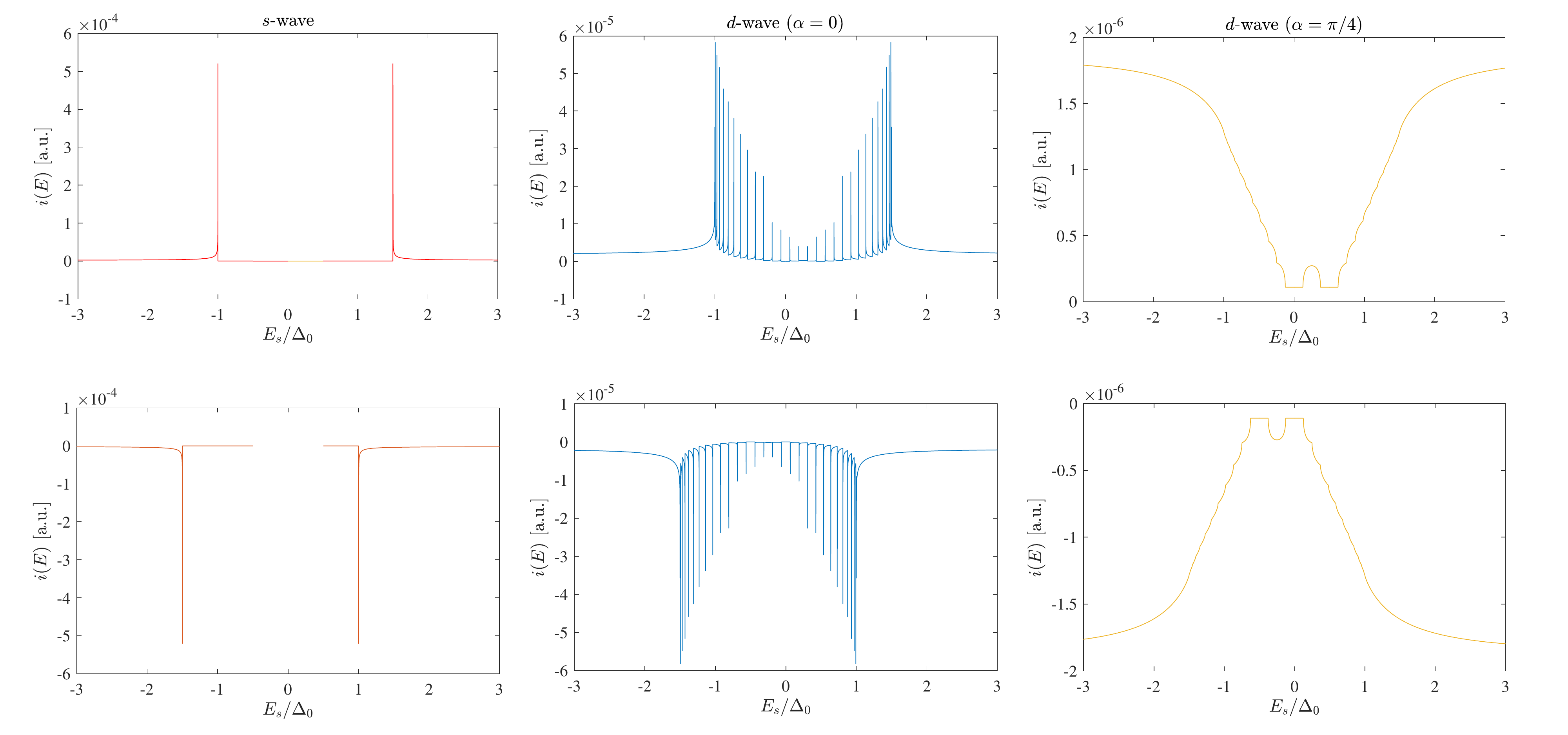}
  \caption{(Color online) Plot of the energy-dependence of the integrands determining the pumped spin current, demonstrating the qualitative difference in spectral spin current for the different superconducting order parameter symmetries. The upper row corresponds to incoming spin$-\uparrow$ electron-like quasiparticles whereas the lower row corresponds to spin$-\downarrow$ electron-like quasiparticles. The $s$-wave case is shown in the left column, $d$-wave $\alpha=0$ in the middle column, $d$-wave $\alpha=\pi/4$ in the right column. Note how the energy-integrands are inverted and shifted by $\hbar\Omega$ for the two spin species. We have used $T/T_c = 0.1$ and $\hbar\Omega/\Delta_0 = 0.5.$
  }%
  \label{fig:intES}
\end{figure*}

\section{Total spin pumping current}

As discussed in the main text, the spin-pumping current induced by an incoming electron-up ($e\uparrow$) particle is calculated by Eq. (\ref{eq:Is}), in which $E_s$ is integrated from $-W_l$ to $W_h$ to ensure the wave vectors in the SC are real and propagating with $E_s^{'}=E_s - \hbar \Omega$. The same integration range applies for spin current induced by $h\downarrow$. As for the spin current based on $e\downarrow$ and $h\uparrow$ incident particles, $E_s$ should be integrated from $-W_h$ to $W_l$ since the energy relation changes to $E_s^{'}=E_s + \hbar \Omega$.

In Eq. (\ref{eq:Is}), the quantum mechanical spin-pumping current $j_{sz}=-\frac{\hbar^2}{2m_e}\Imag[\Psi_S^{\dagger}\partial_x\mathbf{\hat{\sigma_z}}\Psi_S]$ is introduced. %For simplicity, here we consider normal incident (i.e., $\theta = 0$). 
Based on the complete wavefunctions in Appendix A and B, it can be seen that the hole and electron-induced quantum mechanical spin currents fulfill the relation
\begin{equation}
  j_{sz,h\uparrow(\downarrow)}(E_s,\theta) = -j_{sz,e\uparrow(\downarrow)}(-E_s,\pi-\theta). 
\label{eq:j_h_e}
\end{equation}
Let us first consider the resulting spin-pumping current induced by $e\uparrow$ and $h\uparrow$, i.e. 
\begin{equation}
   I_{s,e\uparrow}= \int^{\pi/2}_{-\pi/2} \int_{-W_l}^{W_h} \cos\theta j_{sz,e\uparrow}(E_s,\theta)f_0(E_s)N_sdE_s d\theta,  
\end{equation}
\begin{equation}
   I_{s,h\uparrow}=\int^{\pi/2}_{-\pi/2} \int_{-W_h}^{W_l} \cos\theta j_{sz,h\uparrow}(E_s,\theta)f_0(E_s)N_sdE_s d\theta. 
\label{eq:I_h}
\end{equation}
Inserting Eq. (\ref{eq:j_h_e}) into Eq. (\ref{eq:I_h}) and using $f_0(-E_s)=1-f_0(E_s)$, $j_{sz}(\theta) = j_{sz}(\theta+\pi)$ for any type of incoming particle, and $N_s(E_s)=N_s(-E_s)$, we have
\begin{equation}
  I_{s,e\uparrow}=\frac{1}{2} (I_{s,e\uparrow}+I_{s,h\uparrow}+I_\text{bg}), 
\end{equation}
in which the background contribution $I_\text{bg}= \int^{\pi/2}_{-\pi/2} \int_{-W_l}^{Wh} \cos\theta j_{sz,e\uparrow}(E_s,\theta)N_sdE_s d\theta$. Similar relations apply for incident spin-$\downarrow$ particles. Therefore,  there are two equivalent ways of computing the spin current.
The first way considers only electrons and no hole excitations. The second way considers both electrons and holes, in which case a background spin current contribution $I_\text{bg}$ must also be included. Consequently, we may calculate the total spin current as the sum of electron-up
and electron-down contributions without involving the hole incidents, i.e., $I_{s}=I_{s,e\uparrow}+ I_{s,e\downarrow}$. The appearance of the background contribution can be understood most easily by considering a normal metal with second-quantized Hamiltonian $\hat{H} = \sum_{\vk\sigma} \epsilon_\vk c_{\vk\sigma}^\dag c_{\vk\sigma}$. Here, $\epsilon_{\vk}$ is the energy dispersion relation and $c,c^\dag$ are fermion annihilation/creation operators. Using that $c_{\vk\sigma}^\dag c_{\vk\sigma} = 1 - c_{\vk\sigma} c_{\vk\sigma}^\dag$, we can rewrite the free electron gas form above to a combined electron and hole gas form plus a background contribution consisting of all electron states being filled:
\begin{align}
\hat{H} = \frac{1}{2} \sum_{\vk\sigma} \epsilon_\vk c_{\vk\sigma}^\dag c_{\vk\sigma} - \frac{1}{2} \sum_{\vk\sigma} \epsilon_\vk c_{\vk\sigma} c_{\vk\sigma}^\dag + \frac{1}{2} \sum_{\vk\sigma} \epsilon_\vk.
\label{eq:bg}
\end{align}
The last term on the right hand side of Eq. (\ref{eq:bg}) is consistent with the background spin current term $I_\text{bg}$ due to the absence of a Fermi-Dirac distribution function in its expression.

\section{Energy-dependence of spin current integrand}

It is instructive to consider the energy-resolved quantum mechanical spin current for the $s$-wave and $d$-wave SCs, respectively, and see how the low-energy behavior changes depending on $\alpha$. We define the quantity $i(E_s) = \frac{2m_{e}}{\hbar^2}\int^{\pi/2}_{-\pi/2} j_{sz}(E_s,\theta) N_s \cos\theta d\theta$. In Fig. \ref{fig:intES}, we plot the integrand $i(E_s)$ determining the pumped spin current as a function of energy $E_s$ for different superconducting order parameter symmetries. Comparing the integrands corresponding to $e\uparrow$ and $e\downarrow$ quasiparticles shown in the upper and lower rows, an energy shift of precisely $\hbar\Omega/\Delta_0$ between them can be observed. This energy shift also appears in the different energy integration ranges for $e\uparrow$ (i.e., $[-W_l,W_h]$) and $e\downarrow$ (i.e., $[-W_h,W_l]$) incident particles with $W_h=W_l+\hbar\Omega$. In addition, due to the specific form of the density of states $N_s$ and coherence factors of the SC, it can be seen that the main contributions of the integrands well below $T_c$ are within a small energy range around the gap edge. Meanwhile, the sum of integrands of $e\uparrow$ and $e\downarrow$ approaches zero for large $E_s$ values. Therefore, it is more efficient to use smaller energy integration range to cover the integrated spin current and reduce possible numerical error induced by the extremely sharp peaks appearing in Fig. \ref{fig:intES}. On the other hand, a larger energy integration range should be considered when the temperature approaches $T_c$. For $T>T_c$, the full energy range $[-W_l,W_h]$ for $e\uparrow$ and $[-W_h,W_l]$ for $e\downarrow$ must be considered since the SC then behaves as NM with zero gap.


\begin{thebibliography}{999}

\bibitem{tserkovnyak_prl_02} Y. Tserkovnyak, A. Brataas, and G. E. W. Bauer
Phys. Rev. Lett. \textbf{88}, 117601 (2002)



\bibitem{takahashi_handbook} S. Takahashi, \textit{Physical Principles of Spin Pumping.} In: Xu, Y., Awschalom, D., Nitta, J. (eds) Handbook of Spintronics. Springer, Dordrecht  (2016).

\bibitem{ralph_jmmm_08} D. C. Ralph and M. D. Stiles, J. Mag. Mag. Mater. \textbf{320}, 1190 (2008). 

\bibitem{zutic_rmp_04} I. Zutic, J. Fabian, ans S. Das Sarma, Rev. Mod. Phys. \textbf{76}, 323 (2004). 

\bibitem{kittel_pr_48} C. Kitte, Phys. Rev. \textbf{73}, 115 (1948). 



\bibitem{tserkovnyak_rmp_05} Y. Tserkovnyak, A. Brataas, G. E. W. Bauer, and B. I. Halperin, Rev. Mod. Phys. \textbf{77}, 1375 (2005). 


\bibitem{bell_prl_08} C. Bell, S. Milikisyants, M. Huber, and J. Aarts, Phys.
Rev. Lett. \textbf{100}, 047002 (2008).

\bibitem{morten_epl_08} J. P. Morten, A. Brataas, G. E. W. Bauer, W. Belzig,
and Y. Tserkovnyak, Europhys. Lett. \textbf{84}, 57008 (2008).

\bibitem{jeon_natmat_18} K.-R. Jeon, C. Ciccarelli, A. J. Ferguson, H. Kurebayashi,
L. F. Cohen, X. Montiel, M. Eschrig, J. W. A. Robinson
and M. G. Blamire, Nature Mater. \textbf{17}, 499 (2018).

\bibitem{jeon_prapp_18} K.-R. Jeon, C. Ciccarelli, H. Kurebayashi, J. Wunderlich,
L. F. Cohen, S. Komori, J. W. A. Robinson, and M. G.
Blamire, Phys. Rev. Applied \textbf{10}, 014029 (2018).

\bibitem{yao_prb_18} Y. Yao, Q. Song, Y. Takamura, J. P. Cascales, W. Yuan,
Y. Ma, Y. Yun, X. C. Xie, J. S. Moodera, and W. Han,
Phys. Rev. B \textbf{97}, 224414 (2018).

\bibitem{umeda_apl_18} M. Umeda, Y. Shiomi, T. Kikkawa, T. Niizeki, J.
Lustikova, S. Takahashi, and E. Saitoh, App. Phys. Lett.
\textbf{112}, 232601 (2018).

\bibitem{inoue_prb_17} M. Inoue, M. Ichioka, and H. Adachi, Phys. Rev. B \textbf{96},
024414 (2017).

\bibitem{kato_prb_19} T. Kato, Y. Ohnuma, M. Matsuo, J. Rech, T. Jonckheere, T. Martin, 	Phys. Rev. B \textbf{99}, 144411 (2019)

\bibitem{cheng_prl_14} R. Cheng, J. Xiao, Q. Niu, and A. Brataas,
Phys. Rev. Lett. \textbf{113}, 057601 (2014)

\bibitem{vaidya_science_20} P. Vaidya, S. A. Morley, J. V. Tol, Y. Liu, R. Cheng, A. Brataas, D. Lederman, and E. del Barco, Science \textbf{368}, 160 (2020)

\bibitem{fyhn_prb_21} E. H. Fyhn and J. Linder, Phys. Rev. B \textbf{103}, 134508 (2021).


\bibitem{ominato_prb_22} Y. Ominato, A. Yamakage, T. Kato, M. Matsuo, Phys. Rev. B \textbf{105}, 205406 (2022).

\bibitem{ominato_prb_22a} Y. Ominato, A. Yamakage, and M. Matsuo, Phys. Rev. B \textbf{106}, L161406 (2022).

\bibitem{carreira_prb_21} S. J. Carreira, D. Sanchez-Manzano, M.-W. Yoo, K. Seurre, V. Rouco, A. Sander, J. Santamaria, A. Anane, and J. E. Villegas, Phys. Rev. B \textbf{104}, 144428 (2021).








\bibitem{tanaka_prl_95} Y. Tanaka ans S. Kashiwaya, Phys. Rev. Lett. \textbf{74}, 3451 (1995). 

\bibitem{hu_prl_94} C.-R. Hu, Phys. Rev. Lett. \textbf{72}, 1526 (1994).

\bibitem{kashiwaya_prb_99} S. Kashiwaya, Y. Tanaka, N. Yoshida, and M. R. Beasley, Phys. Rev. B \textbf{60}, 3572 (1999).









\bibitem{eremin_prb_06} I. Eremin, F. S. Nogueira, and R.-J. Tarento
Phys. Rev. B \textbf{73}, 054507 (2006).

\bibitem{gronsleth_prl_06} M. S. Gr{\o}nsleth, J. Linder, J.-M. Børven, and A. Sudb{\o}
Phys. Rev. Lett. \textbf{97}, 147002 (2006)

\bibitem{brydon_prb_09} P. M. R. Brydon
Phys. Rev. B 80, 224520 – Published 22 December 2009; Erratum Phys. Rev. B 85, 099905 (2012)

\bibitem{tanaka_prb_09} Y. Tanaka, T. Yokoyama, A. V. Balatsky, and N. Nagaosa
Phys. Rev. B \textbf{79}, 060505(R) (2009).

\bibitem{takashima_prb_17} R. Takashima, S. Fujimoto, and T. Yokoyama,
Phys. Rev. B \textbf{96}, 121203(R) (2017)

\bibitem{johnsen_prl_21} L. G. Johnsen, H. T. Simensen, A. Brataas, and J. Linder, Phsy. Rev. Lett. \textbf{127}, 207001 (2021). 


\bibitem{btk} G. E. Blonder, M. Tinkham, and T. M. Klapwijk, Phys. Rev. B \textbf{25}, 4515 (1982). 

\bibitem{ahari_prb_21} M. T. Ahari and Y. Tserkovnyak, Phys. Rev. B \textbf{103}, L100406 (2021).

\end{thebibliography}
\end{document}